\documentclass[ reprint, amsmath,amssymb,prb,aps]{revtex4-2}
\usepackage{color}
\usepackage{verbatim}
\usepackage{lipsum}
\usepackage{hyphenat}
\usepackage{dirtytalk}
\usepackage{graphicx}
\usepackage{dcolumn}
\usepackage{bm}
\usepackage{tabularx}
\usepackage{hyperref}
\usepackage{mathtools}
\usepackage{subfigure}
\usepackage{tabularx}
\usepackage{etoolbox}
\usepackage{colortbl}
\preprint{APS/123-QED}
\newsavebox{\measurebox}
\newcolumntype{l}{X}
\newcolumntype{s}{>{\hsize=0.3\hsize}X}
\newcolumntype{m}{>{\hsize=0.85\hsize}X}

\begin{document}
\title{Sharp electromagnetically induced absorption via balanced interferometric excitation in a microwave resonator}

\author{Michael T. Hatzon}
\affiliation{Quantum Technologies and Dark Matter Labs, Department of Physics, University of Western Australia, 35 Stirling Highway, Crawley, WA 6009, Australia.}
\email{22873723@student.uwa.edu.au,  michael.tobar@uwa.edu.au}
\author{Graeme R. Flower}
\affiliation{Quantum Technologies and Dark Matter Labs, Department of Physics, University of Western Australia, 35 Stirling Highway, Crawley, WA 6009, Australia.}
\author{Maxim Goryachev}
\affiliation{Quantum Technologies and Dark Matter Labs, Department of Physics, University of Western Australia, 35 Stirling Highway, Crawley, WA 6009, Australia.}
\author{Jeremy F. Bourhill}
\affiliation{Quantum Technologies and Dark Matter Labs, Department of Physics, University of Western Australia, 35 Stirling Highway, Crawley, WA 6009, Australia.}
\author{Michael E. Tobar}
\affiliation{Quantum Technologies and Dark Matter Labs, Department of Physics, University of Western Australia, 35 Stirling Highway, Crawley, WA 6009, Australia.}
\email{michael.tobar@uwa.edu.au}

\date{\today}

\begin{abstract}
A cylindrical TM$_{0,1,0}$ mode microwave cavity resonator was excited using a balanced interferometric configuration that allowed manipulation of the electric field and potential within the resonator by adjusting the phase and amplitude of the interferometer arms driving the resonator. With precise tuning of the phase and amplitude, 25 dB suppression of the electric field at the resonance frequency was achieved while simultaneously resonantly enhancing the time-varying electric-scalar potential. Under these conditions, the system demonstrated electromagnetically induced absorption in the cavity response due to the annulment of the electric field at the resonance frequency. This phenomena can be regarded as a form of extreme dispersion, and led to a sharp increase in the cavity phase versus frequency response by an order of magnitude when compared to the cavity Q-factor. This work presents an experimental setup that will allow the electric-scalar Aharonov-Bohm effect to be tested under conditions involving a time-varying electric-scalar potential, without the presence of an electric field or magnetic vector potential, an experiment that has not yet been realised.
\end{abstract}
\pacs{}
\maketitle

\section{Introduction}\label{intro}

In classical physics, electromagnetic fields are considered physical observables \cite{kaloyerou1994casual}, with field measurements being a standard procedure in many applications. These fields exert forces on charges through the Lorentz force. Alternatively, potentials like the electric-scalar potential $V$ and the magnetic-vector potential $\vec{A}$ offer a different way to describe classical electrodynamics \cite{griffiths2021introduction}, though their physical influence on charged particles is less straightforward than that of the fields. In 1959, Aharonov and Bohm (AB) published a paper demonstrating that potentials can interact with charges without the direct influence of electromagnetic fields, leading to an accumulated phase (now known as a Berry phase). This effect, often observed as the phase shift of an electron wavefunction, has been confirmed and measured through interactions with quantum dots for example \cite{schuster1997phase}. Additional theoretical support \cite{aharonov1959significance} and experimental evidence \cite{phatak2010three} strongly validate the interaction of charge with the magnetic-vector potential. There has also been an experiment that measured charge interactions involving both the scalar-electric and magnetic-vector potentials, although it occurred in the presence of fields \cite{van1998magneto}.

The AB effect \cite{aharonov1959significance} suggests that in quantum mechanics, scalar and vector potentials hold a more significant physical role than what classical electromagnetism implies. In classical physics, a particle's motion can be entirely described by fields, as the Lorentz force relies on fields, charges, and currents. In contrast, quantum mechanics requires a more rigorous approach through the canonical formalism, which necessitates the inclusion of potentials in the fundamental equations of motion. This incorporation has been demonstrated to lead to effects such as shifts in interference fringes in particle beams \cite{aharonov1961further,ehrenberg1949refractive} and the splitting of energy levels in quantum systems subjected to time-varying potentials \cite{chiao2023energy,chiao2024gravitational,Tobar2024}. It is anticipated that the energy levels of a quantum system will exhibit splitting when exposed to a single time-varying electric-scalar potential, similar to the analysis used in the AC Stark effect caused by a time varying electric field \cite{autler1955stark,delone1999ac,chiao2023energy}.

To date, no experiment has definitively demonstrated the AB effect solely arising from the electric-scalar potential in the absence of a magnetic vector potential and electromagnetic fields. However, modern experimental techniques, as described in \cite{chiao2023energy}, have proposed a device specifically designed to measure this effect using a Faraday shield that effectively screens the electric field while maintaining a strong electromagnetic potential. In this work, we carefully consider the frequency and thickness of the shield, taking into account the skin effect during its design. We introduce an innovative concept of a resonant Faraday structure based on a microwave resonator excited in an interferometric way, so that the electric field is annulled on resonance leading to Electromagnetically Induced Absorption (EIA) \cite{EIA99,Brazhnikov:19,Shakthi2020,DONG2022413936}, a form of extreme dispersion, reminiscent of Electromagnetically Induced Transparency (EIT) \cite{marangos1998electromagnetically, fleischhauer2005electromagnetically}. We have constructed and characterised the first device of this type and unequivocally show at the resonance frequency, when the interferometric excitation is balanced properly so absorption occurs on resonance, a time-varying potential exists within the resonator in the absence of an electric field.

EIA and EIT are processes in which a medium exhibits induced absorption or transparency respectively, over a narrow spectral range with enhanced dispersion \cite{boller1991observation,EIA99}. Similar phenomenon has been observed with atoms in optical cavities \cite{mucke2010electromagnetically} and in the microwave domain using Brillouin scattering \cite{Shakthi2020}. EIT mechanisms can result in low group velocities of radiation, also known as “slow” light, which has a range of quantum applications, such as quantum storage and switching \cite{krauss2008we}, in various media \cite{khurgin2010slow}, including photonic crystals \cite{baba2008slow}. 

In some case EIT/EIA effects may be observed between two resonators as one mode tunes through the bandwidth of the lower-Q mode. In this case the higher-Q mode takes on a Fano structure \cite{fano1961effects} as it tunes through the bandwidth of the lower-Q mode. This effect was shown in the microwave region in 1991 \cite{tobar1991generalized} and was due to a mutual resistance between the two modes. Later the same effect was shown in the optical domain \cite{Li2011,Peng:2014aa,Akhtar:24}. This effect we observe here is different and necessarily involves three cavities. It occurs due to excitation through to the two thin foils, which acts as a mutual resistive coupling from the input cavities to the output cavity. All resonators must be tuned, and exhibit no signs of mode splitting through the foils proving that the excitation is through a resistive rather than reactive coupling. We show that the effect only appears as an absorption with no Fano resonance structure observed with detuning, so is very different to what has been achieved previously.

\section{Device to Test Electric-Scalar AB Effect}\label{scalarab}

\begin{figure}[t]
\centering
\includegraphics[angle=-90,width=1.0\linewidth]{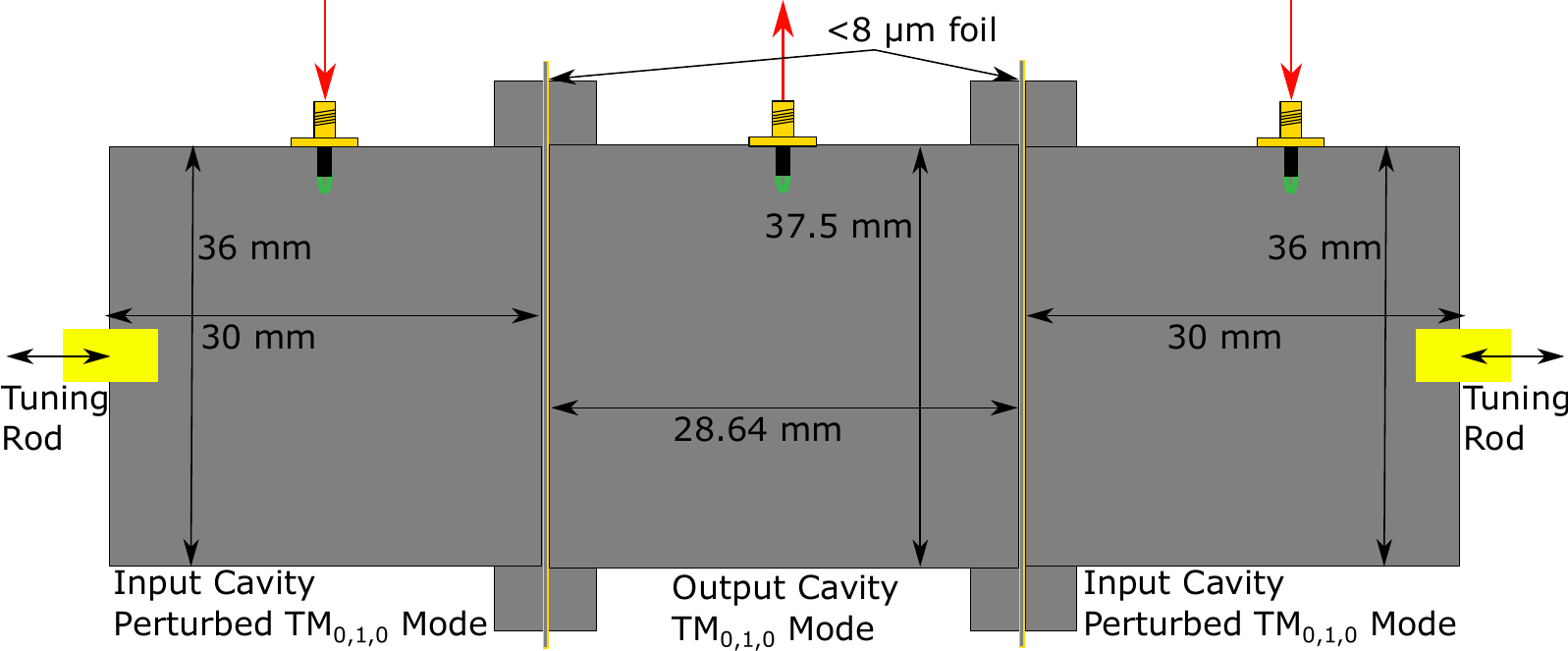}
\caption{Schematic of the multiple microwave cavity resonator system used to create a time varying potential with a suppressed electric field on resonance in the output cavity. Two thin copper foils are bolted between the three sections, to create two input resonators and an output resonator in the middle, which operates in the TM$_{0,1,0}$ mode. Tuning rods are inserted into each of the outer input cavity resonators to create perturbed TM$_{0,1,0}$ modes, which are tuneable. The resonant frequencies are controlled by a micrometer, allowing them all to be tuned to the same value. The electric fields from the input cavities induce charges within the foil.}
\label{ABcavity}
\end{figure}
\begin{figure}[h]
\begin{minipage}{0.5\columnwidth}
\centering
\includegraphics[angle=-90,width=1\columnwidth]{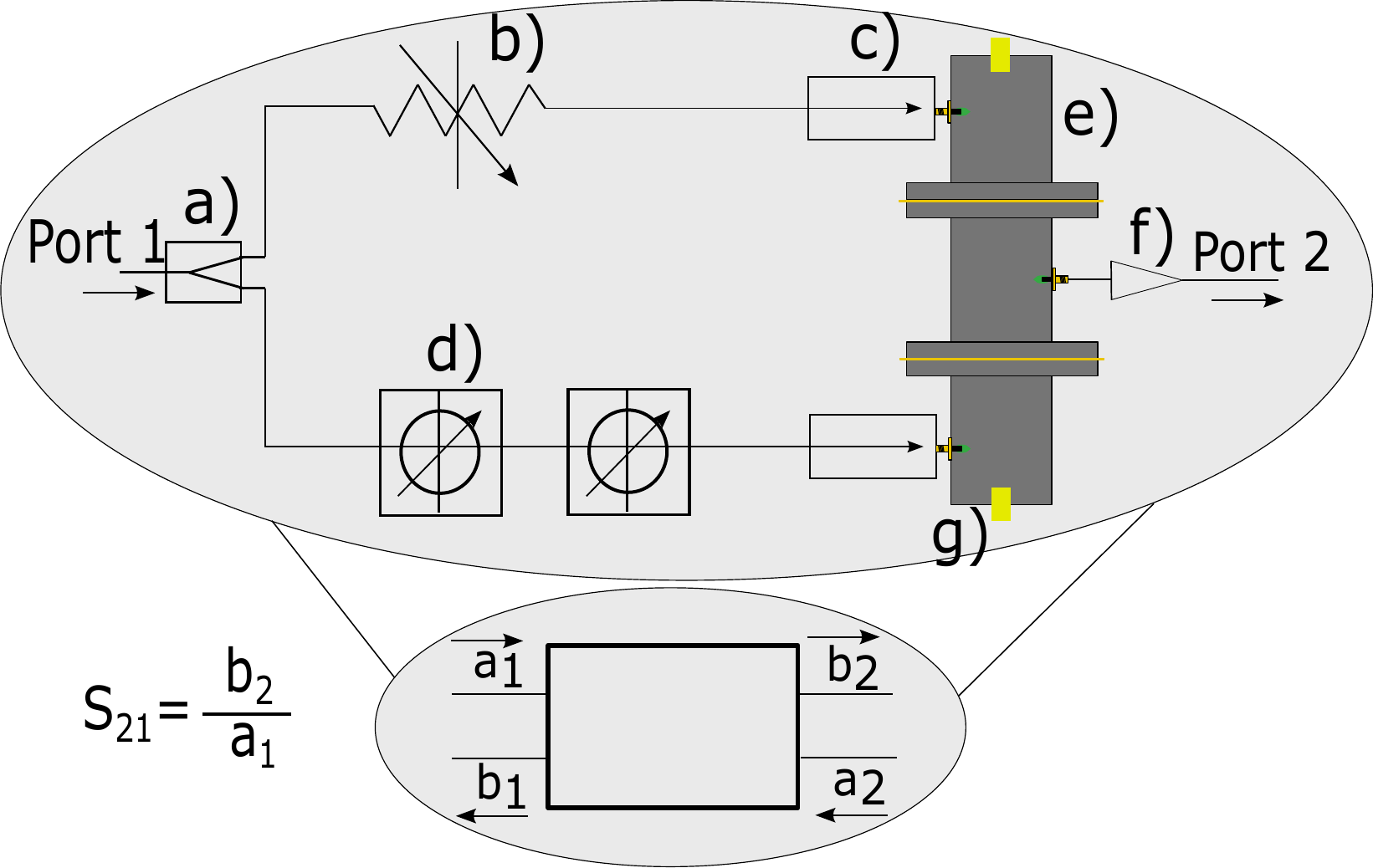}
\end{minipage}%
\begin{minipage}{0.5\columnwidth}
\includegraphics[width=1\columnwidth]{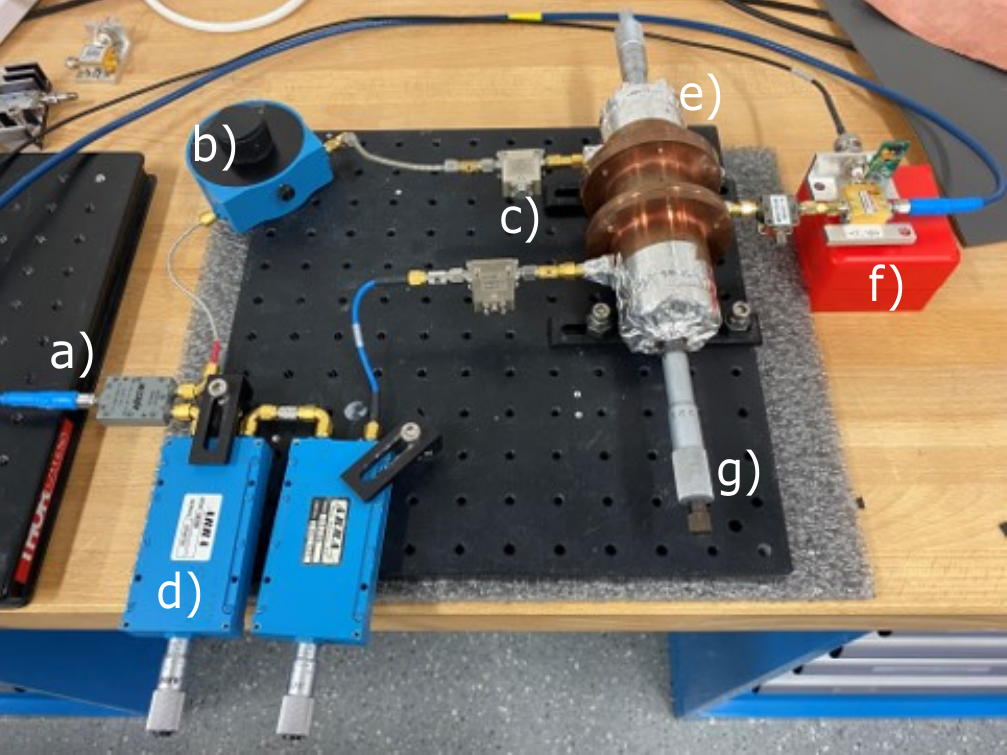}
\end{minipage}
\caption{Left, a schematic illustrating the AB circuit components: a) power splitter, b) variable attenuator, c) isolator, d) phase shifter, e) cavity resonator, f) low noise amplifier and g) frequency tuning rods for the input cavities. The setup measures the transmission from port 1 to port 2 ($S_{21}$). Right, a photo of the experimental setup.}
\label{ABsetup}
\end{figure}

To experimentally test the electric-scalar AB effect, it is essential to develop a device capable of generating a time-varying potential without the presence of an electric field. In this work, we achieve this using a multiple cavity structure as shown in Fig.~\ref{ABcavity}. The device utilises a novel interferometric technique of exciting a cavity through two tuneable resonant input cavities, with the output cavity sandwiched between the two input cavities. The cavities are  separated by thin copper foils, so the fields from the input cavities couple into the output cavity through induced charges in the foils. Total control of the electric field and electric-scalar potential can be achieved by controlling the phase and amplitude in the two input cavities, so when tuned and balanced appropriately the output cavity may be excited with a time varying electric potential with suppressed resonant electric field. Finite element software was utilised to predict the electric charge and potential field distributions inside the cylindrical output resonator. To run a test of the electric-scalar AB effect, on would measure the evolution of a quantum system placed in this output cavity.

The device was excited in a manner analogous to a Mach-Zehnder interferometer \cite{hatzon2024microwave}, as depicted in Fig.~\ref{ABsetup}. This configuration allows precise control over the relative phase and amplitude of the electromagnetic field fed into one input cavity relative to the other. Consequently, this enabled the control over the relative phase and amplitude of the charges oscillating on the inner surfaces of both copper foils, facilitating controlled excitation of the output resonator. As $\vec{E}=-\vec{\nabla} V$, when the charges on the two foils are out of phase, a strong spatial gradient of the electric potential is established within the output resonator, leading to maximum excitation via the associated electric field. Conversely, when the charges are in phase and balanced, such that the charges on both foils have the same magnitude and phase, the spatial gradient of the electric potential is suppressed, and thus, the electric field within the output resonator is also suppressed. However, the electric potential remains present within the resonator and is not screened. Under this balanced condition, the electric field is reduced to zero at the resonance frequency, causing EIA on resonance, a form of extreme dispersion, which serves as a verification technique that the balance condition has been met.

\section{Cavity Electric Field and Potential}

In this section, we demonstrate through simulation how to create a suppressed electric field with a time-varying potential inside the output cavity, as depicted in Fig.~\ref{ABcavity}. The first step involves ensuring one excites the output cavity in the $TM_{0,1,0}$ mode, which functions similarly to a high-frequency capacitor. For proper excitation by the input cavities, they too must operate in the TM$_{0,1,0}$ mode and be tuned to the same frequency as the output cavity. However, the frequency of the TM$_{0,1,0}$ mode, given by $f{_{TM_{0,1,0}}} = \frac{c\chi_{0,1}}{R}$, is tuneable only by adjusting the cavity radius ($R$), not its length, making frequency tuning particularly challenging. In this expression, $c$ represents the speed of light, and $\chi_{0,1} = 2.405$ is the first zero of the Bessel function of the first kind, $J_0(\chi)$. For the output cavity with a diameter of 37.5 mm, as shown in Fig.~\ref{ABcavity}, the calculated mode frequency is 6.10 GHz.

\begin{figure}[t]
\centering
\includegraphics[width=1\linewidth]{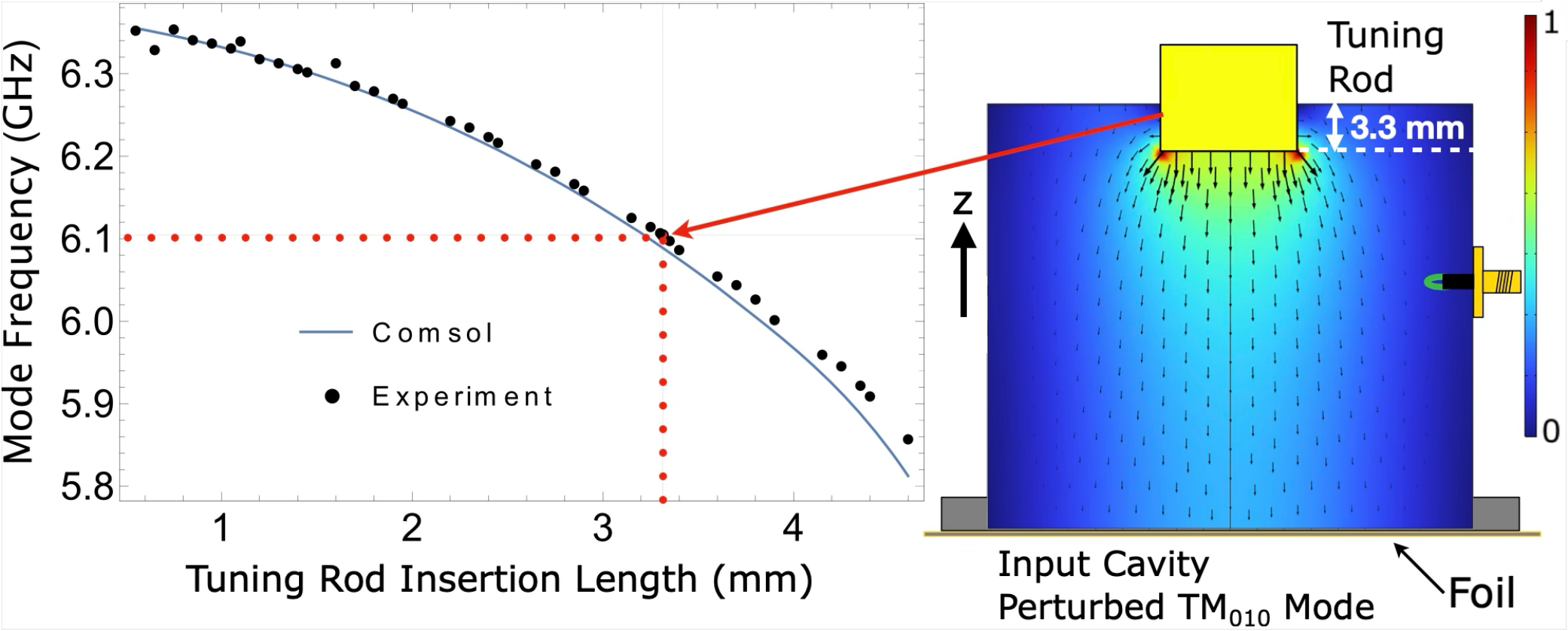}
\caption{Left: Tuning plot for the input cavities, showing experimental results of one of the cavities compared to COMSOL finite element modelling, showing excellent agreement. A rod insertion length of 3.3 mm was required to achieve mode frequency tuning to match that of the central inner output cavity of 6.1 GHz, with the frequency verified using COMSOL (indicated with red dashed lines). Right: The density plot in the $r$-$z$ plane of the magnitude of the electric field of the perturbed TM$_{0,1,0}$ mode field structure caused by the tuning rod, arrows are also include, which represent the vector direction and magnitude of the electric filed. The mode was excited by a loop probe, which coupled to the tangential  $H_{\varphi}$ magnetic field component. The fields near the foil are dominated by the axial z-component of the electric field, similar to the unperturbed fundamental mode TM$_{0,1,0}$ mode.}
\label{perturbed}
\end{figure}

To address the tuning challenge of the input cavities, a tuning rod was inserted, as shown in Fig.~\ref{ABcavity}, to perturb the associated modes. This modification lowered the frequency of the TM$_{0,1,0}$ mode and, if the post was inserted too far, it could eventually transform into a re-entrant cavity mode, as discussed in \cite{LeFloch2013,McAllister2017}. Consequently, both input cavities were designed with a slightly smaller diameter of 36 mm, 1.5 mm less than the output cavity, resulting in a resonant frequency of 6.37 GHz for the $TM_{0,1,0}$ mode when the tuning rod was completely removed. As illustrated in Fig.~\ref{perturbed}, to tune the input cavities resonance frequencies to the output cavity, required the tuning rod to be inserted by 3.3 mm. Despite the perturbation of the input cavity modes, they still exhibit a strong electric field oriented along the cylindrical z-axis, terminating on the foils and thus inducing charge and current distributions of similar form to the unperturbed distribution. These current and charges excite the output cavity mode, however they will be attenuated by the skin effect within the copper foils. For example, the current density through the conductor thickness will be of the form  $|\vec{J}(z)| = |\vec{J}(0)|e^{-\kappa z}$, where  $\kappa^{-1}$ is the skin depth \cite{griffiths2021introduction,1130282272498957568,wheeler1942formulas}. The critical aspect of this design is to ensure the foils are thick enough to reflect the majority of the electromagnetic field back into the cavities, but not too thick as to prevent sufficient leakage of current, charge, and electric field to the opposite side of the foil, necessary to excite the output cavity mode in a similar way to a cavity probe. For this aim we implemented a $d=8~\mu$m thick copper foil, as shown in Fig.~\ref{ABcavity}. Given that the skin depth of copper is approximately 0.8 $\mu m$ at 6.10 GHz, the $10$ skin-depth thick foil therefore reduces current density amplitudes by a factor of $4.5\times10^{-5}$ when $z=d$. This reduction is nonetheless sufficient to excite modes in the output cavity, as demonstrated by our experiments. By comparison, increasing the foil thickness to $d=50~\mu$m, which corresponded to approximately 62.5 skin depths and an attenuation of the current density by a factor of $7\times10^{-28}$, the output mode resonance could not be measured.

\begin{figure}[t]
\centering
\includegraphics[width=1\linewidth]{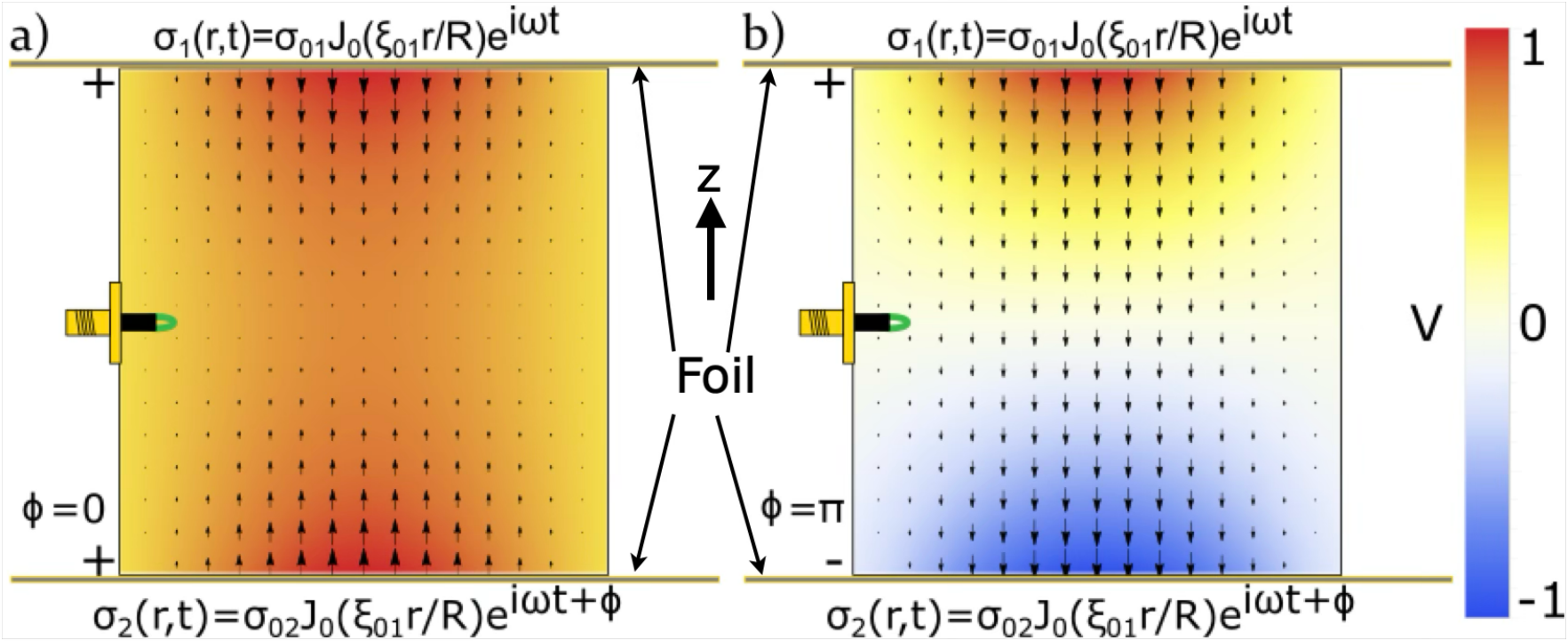}
\caption{A cross-section view in the $r$-$z$ plane of the central output cavity (z-axis shown), with the calculated electric-scalar potential distribution, V (color scale on the right), and z-component of the normalised electric field, $E_{z}$ (arrows with length indicating magnitude) with $|E_{0}|=1$. These arise from the analytical solutions to cavity TM$_{0,1,0}$ electric field modes leaking through a thin foil (\textless8) $\mu$m above and below the central cavity. Here, we consider the charge is balanced on the foils so $\sigma_{01}=\sigma_{02}$, and vary $\phi$. a) The potential and field distributions when $\phi=0$ resulting in large electric-scalar potential and suppressed electric field at the cavity centre. b) The potential and field distributions when $\phi=\pi$ creating suppressed electric-scalar potential and large electric field at the cavity centre. The resonantor was excited by a loop probe, which coupled to the tangential  $H_{\varphi}$ magnetic field component as shown.}
\label{ABcentralfield}
\end{figure}

Assuming the input cavities are driven at a frequency of $\omega$, the standard form of the electric field vector-phasor, for the output cavity TM$_{0,1,0}$ mode, will be given by,
\begin{equation}
\tilde{\vec{E}}(r,t)=\tilde{E_{0}} \begin{bmatrix}
           0 \\
           0 \\
           J_{0}(\frac{\chi_{0,1}}{R}r)
          \end{bmatrix}e^{i \omega t},
          \label{tmfield1}
\end{equation} 
which depends on the radial coordinate $r$, and the cavity radius, $R$. The corresponding surface charge distribution, $\sigma(r,t)$, that accumulates on the inner surfaces of the foils produced by the electric field can be calculated from the following surface integral,
\begin{equation}
\int \vec{E}.d\vec{S}=\frac{1}{\epsilon_{0}}\int \sigma dS,
\end{equation}
where $\epsilon_{0}$ is the permittivity of free space. This results in $\tilde{\sigma}(r,t)=\epsilon_{0}\tilde{E}_z(r,t)$, with the magnitude of the charge density phasor on the two foils, labeled $1$ and $2$ and shown in Fig. \ref{ABcentralfield}, given by,
\begin{equation}
\tilde{\sigma}_{1,2}(r, t)=\sigma_{01,2} J_0\left(\chi_{01} r / R\right) e^{i\omega t+\phi_{1,2}},
\end{equation}
assuming they are excited with independent amplitudes, $\sigma_{01,2}$, and phases, $\phi_{1,2}$. Subsequently, the electric-scalar potential phasor, $\tilde{V}(\vec{r}) e^{i \omega t}$, can be determined using the following surface integral over the foil,
\begin{equation}
V(\vec{r})=\frac{1}{4 \pi \epsilon_{0}}\int \frac{\sigma(\vec{r} ')}{|\vec{r}-\vec{r}'|}dS,
\end{equation} 
where $\vec{r}$ is the vector from the origin to any point within the resonator and $\vec{r}'$ is from the origin to any point from the inner surface of the foil. Following this the electric field within the resonator may be calculated from, $\vec{E}=-\vec{\nabla}V$.

\begin{figure}[h]
\centering
\includegraphics[width=0.6\linewidth]{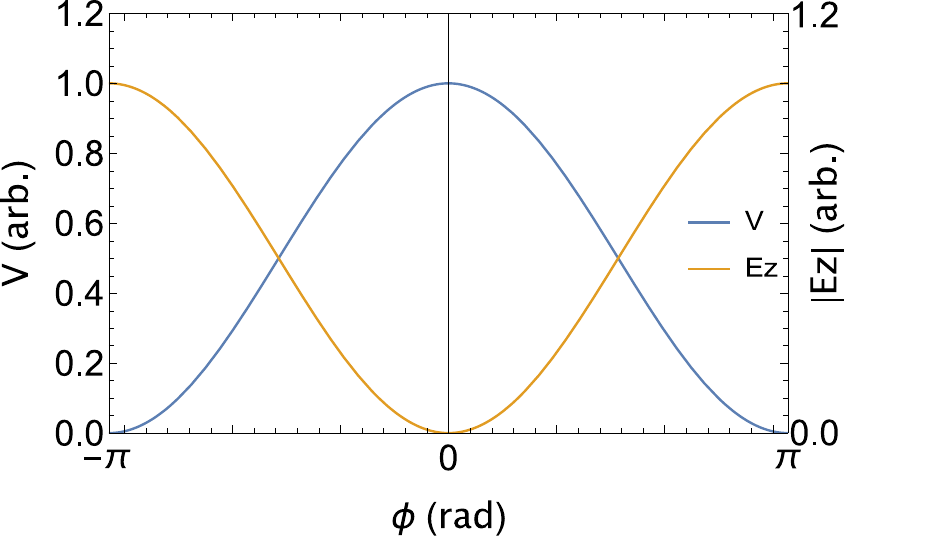}
\caption{The variation of the electric-scalar potential, $V$, and the magnitude of the z-component of the electric field, $|E_z|$, at the center of the resonator as a function of interferometer phase shift $\phi$. The electric-scalar potential is maximised when the $E_{z}$ field is minimised, producing the ideal conditions to test the electric-scalar AB effect. }
\label{EzVcomparison}
\end{figure}

Fig.~\ref{ABcentralfield} shows the electric field and potential distribution within the cavity resonator when $\sigma_{01}=\sigma_{02}$, and $\tilde{\sigma}_{1}$ and $\tilde{\sigma}_{2}$ are in phase ($\phi=0$) and out of phase ($\phi=\pi$). When the charges on the foils are in phase, there is a minimum potential gradient, and hence minimum electric field, whilst maintaining a strong potential, making the centre of the cavity an ideal location for testing the electric-scalar AB effect. Fig.~\ref{EzVcomparison} shows the potential, $V$ and the electric field component, $E_z$ in the centre of the cavity as a function of phase assuming $\sigma_{01}=\sigma_{02}$.

\section{Interferometric Excitation: Experimental Results}

In this section, we present the experimental results and verify that the device enables a time-varying potential with a suppressed electric field. The experimental interferometric setup was configured as shown in Fig.~\ref{ABsetup}. To establish a suppressed electric field configuration in the central microwave cavity resonator, equal magnitudes of charge must be effectively deposited on both foils. Achieving balanced charge distribution,  $\sigma_{01}=\sigma_{02}$, requires the circulating power in the two outer resonators to be matched, which is accomplished by incorporating a variable attenuator in one of the interferometer arms. Additionally, the phase difference of the driving microwave signals, arising from variations in the phase length of the interferometric arms, must be carefully aligned using a phase shifter in one of the arms to synchronise the charge and set $\phi=0$. This setup provides complete control over both phase and amplitude of the two excitation charge densities, making it ideally suited for conducting this experiment.

\subsection{Determining Electric Field Suppression}

\begin{figure}[t]
\centering
\includegraphics[width=1.0\linewidth]{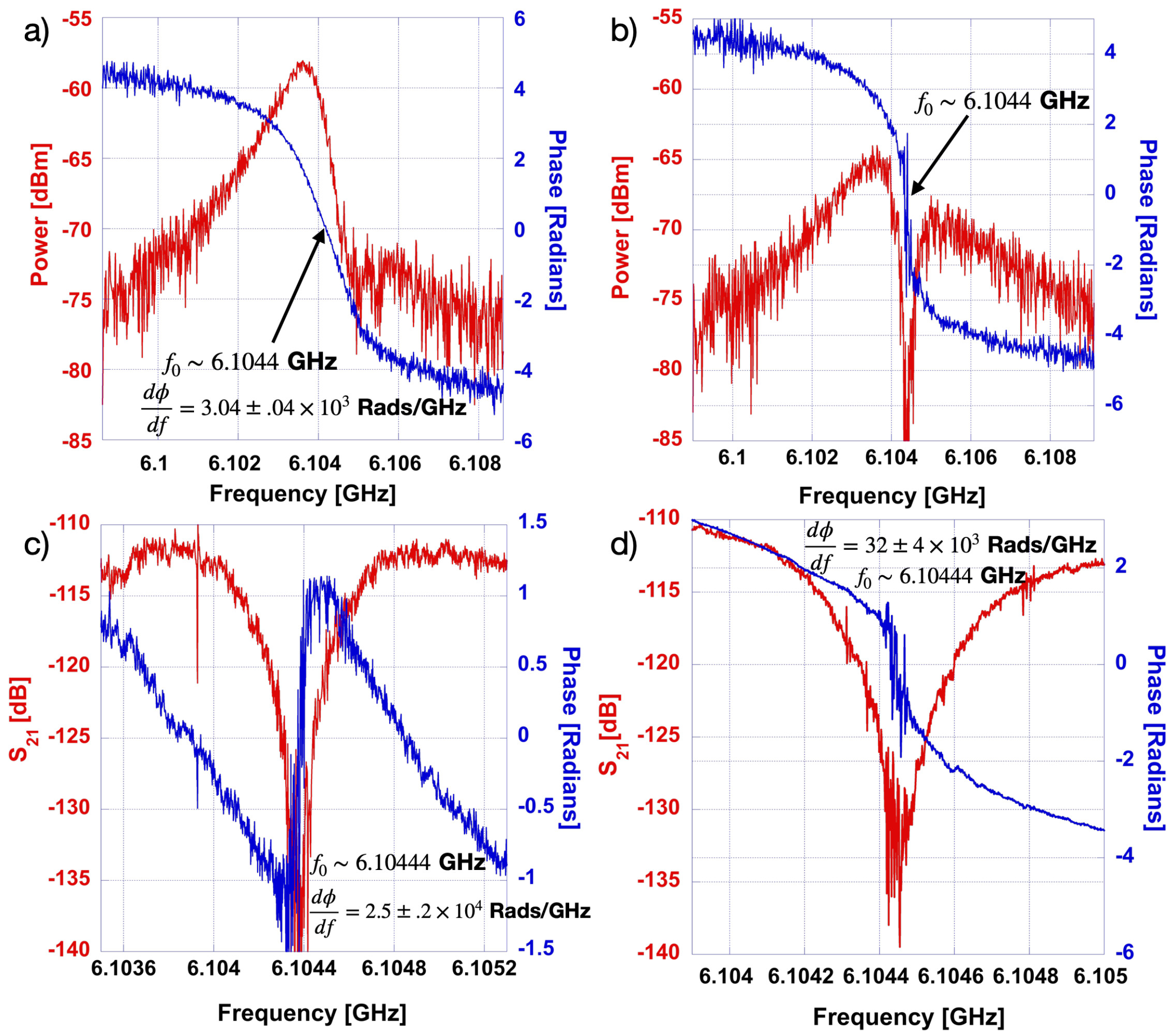}
\caption{Amplitude and phase response in dBm and radians respectively of the balanced interferometer experiment shown in  Fig.~\ref{ABsetup}. The experiment uses a 30 dB amplifier at the output position (port 2) and 20 dBm input power from the VNA at port 1. The responses are measured at port 2 with respect to the output of the amplifier, with the  phase of the interferometer set, a) out of phase $\phi=\pi$, and b) in phase $\phi=0$, the latter resulting in the electric field suppression at the resonance frequency indicated by the arrow of about 25 dB. c) Close up of the $S_{21}$ transmission spectra and phase response, showing the EIA regime with the suppressed electric field causing extreme dispersion, the phase response in this case resembles that of an under coupled resonator. d) Includes a 30 dB high power amplifier at the input with 0 dBm input (instead of 20 dBm) to improve the signal-to-noise ratio, allowing further optimisation of the phase response, through tweaking the balance conditions. An order of magnitude improvement in phase sensitivity was achieved at the higher power than the phase response in a). In the extreme dispersion regime the interferometer became very sensitive to external vibrations causing the error in the fits to the phase response to increase.}
\label{phasetune}
\end{figure}

\begin{figure}[t]
\centering
\includegraphics[width=0.6\linewidth]{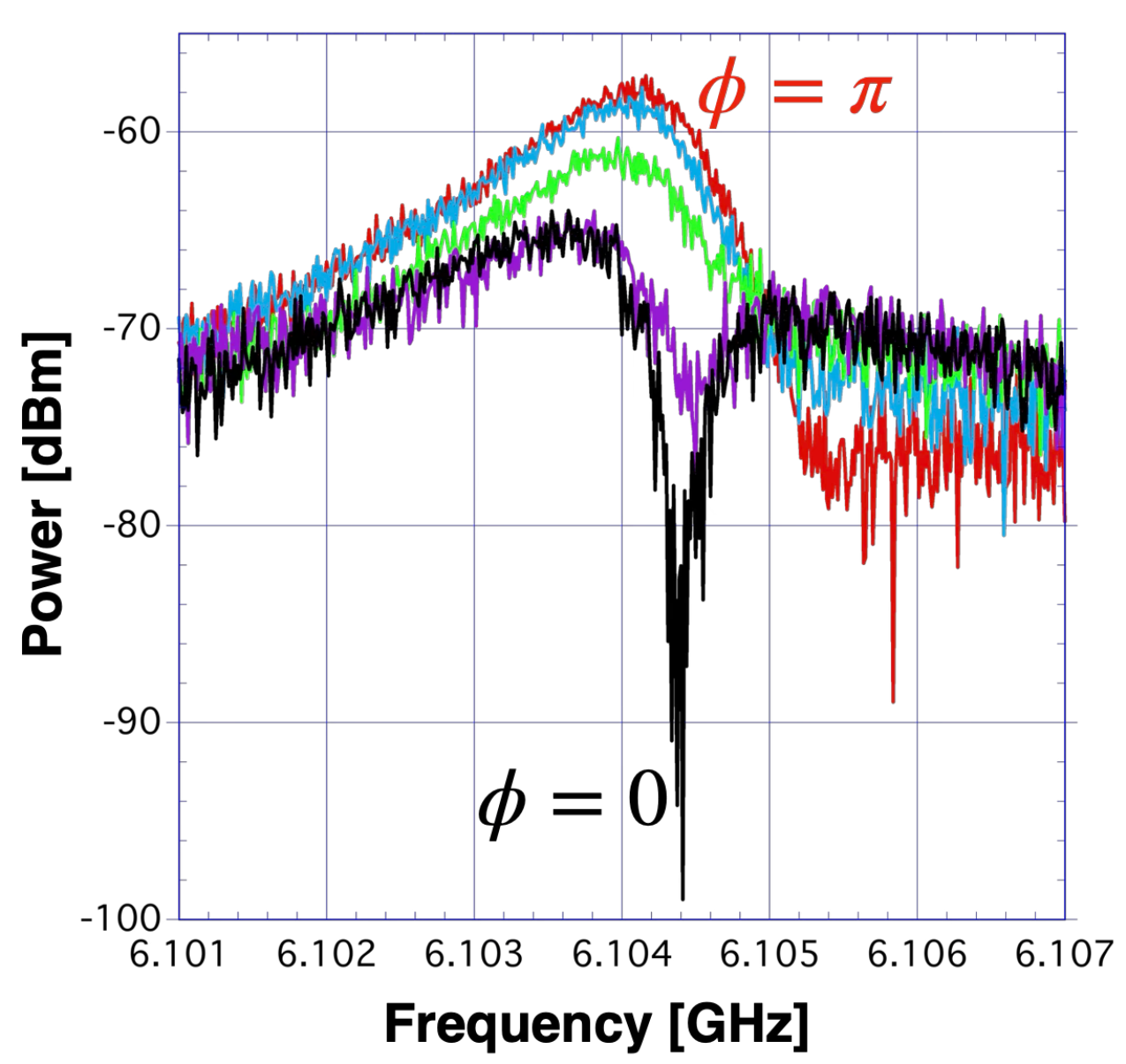}
\caption{Amplitude-frequency response in dBm, for the balanced interferometer when the relative phase between the interferometer arms is varied from, in-phase $\phi=0$ (black) to out of phase, $\phi=\pi$ (red). The purple, green and cyan curves, show increasing intermediate phases, showing reduced suppression and smaller value of electromagnetic induce absorption on resonance.}
\label{PhaseVary}
\end{figure}

When all three resonator frequencies were precisely tuned and driven with the interferometric setup of Fig.~\ref{ABsetup}, we observed that the frequency dependence of the power output from the central cavity was resonant, with an example spectra shown in Fig.~\ref{phasetune}a. This indicates that the outer cavities are capable of exciting the TM$_{0,1,0}$ mode of the central cavity by coupling through the foil boundaries. The specific case of Fig.~\ref{phasetune}a was when the electric field amplitudes inside the input cavities were equal and opposite ($\phi=\pi$). Note, that the cavity response was not quite Lorentzian, which might be expected since there are three resonators involved in the transmission through the interferometer and appeared to contain some Fano distortion. For equal amplitude driving fields, the field suppression effect, shown in Fig.~\ref{phasetune}b occured when the two input cavities were oscillating in phase ($\phi=0$) with one another. This resulted in the nullification of the mode's transmission of the output resonance within a narrow microwave region at the resonance frequency of 6.1044 GHz, leading to extreme dispersion. A narrower span of this effect is depicted in Fig. \ref{phasetune}c and d, the spectra exhibit approximately 25 dB electric field suppression. These suppression levels could be further enhanced with finer control of the attenuation, cavity frequencies, and phase shift, as well as by increasing the output amplification to improve the signal-to-noise ratio for the VNA trace. We also noted the significant increase in phase response in the extreme dispersion regime, which could be advantageous for other applications, which require phase or frequency control. For example, Fig. \ref{phasetune}a shows a measured phase sensitivity of $\frac{d\phi}{df}=(3.04\pm0.04)\times10^3$ GHz/Rad, while Fig. \ref{phasetune}d shows a measured phase sensitivity of $\frac{d\phi}{df}=(32\pm4)\times10^3$ GHz/Rad, a factor of 10.5 times larger. The device’s interferometric nature introduced a high level of sensitivity, underscored by the small transmission spikes visible in the spectra often corresponding to sudden changes in the laboratory environment, such as vibrations from a door opening. Fig. \ref{PhaseVary} shows the situation where the charge on the foils is balance and the phase is varied, which shows that intermediate phases partially cancel the electric field.

To precisely determine the electric-scalar potential within the microwave cavity, one must first calculate the charge deposited on each foil in the resonant-peak scenario depicted in Fig.~\ref{phasetune}a. The magnitude of this charge will be identical to that in the anti-resonance case, except that both foils will carry the same sign of charge. To calculate the charge magnitude, we first calculate the energy stored on resonance. This may be determined from the first-principles definition of the Q-factor (Q=Energy stored/power dissipated per cycle), the loaded form of which may be estimated from the measured phase response given by, 
\begin{equation}
Q_{L}=\frac{f_0}{2}\frac{d\phi}{df}|_{f=f_0},
\label{PhaseResponse}
\end{equation}
the well known phase response of a Lorentzian at the resonance frequency, $f_0$. Thus from the measured phase response shown in Fig. \ref{phasetune}a, the loaded Q-factor was determined to be $Q_L=(9.28\pm.12)\times10^3$. Following this, the stored energy, $U_{st}$, may be calculated from \cite{HARTNETT2011}:
\begin{equation}
U_{st}=\frac{P_{out} Q_{L}(1+\beta)}{2\pi f_0 \beta},
\end{equation} 
where $P_{out}$ is the output power from the resonator before amplification and $\beta$ is the coupling of the central cavity probe. With the output coupling set as $\beta=1.0$ from reflection measurements, and $P_{out}=-87.6$ dBm (from the maximum amplitude of Fig.~\ref{phasetune}a less 30 dB amplification), we calculate $U_{st}=8.4\times10^{-19}$ J. The electric field for the TM$_{0,1,0}$ mode may then be determined by relating this to the energy stored in electromagnetic fields, to obtain,
\begin{equation}
U_{st}=\frac{\epsilon_{0}}{2}\int_{V}(\tilde{\vec{E}}.\tilde{\vec{E^{*}}})dV =\frac{\epsilon_{0}}{2}|E_{0}|^{2}\int_{V}(J_{0}(\frac{\chi_{0,1}}{R}r))^2dV.
\label{pstored}
\end{equation}
Solving for $|E_{0}|$ in Eq.~\ref{pstored} by using the calculated value of stored energy, $U_{st}$ we obtain $|E_{0}|\sim0.15~V/m$, and then using the relation for surface charge density, 
\begin{equation}
\sigma(r)=\epsilon_{0}E_{z},
\end{equation} 
we determined the charge density amplitudes to be, $|\sigma_{01,2}|\sim1.3\times10^{-12} C/m^2$. Thus, we calculated an electric-scalar potential amplitude of Fig.~\ref{ABcentralfield}, to be $|V_{0}|\sim0.66$ mV at the centre of the output cavity, and we predict that the energy level splitting due to the electric-scalar AB effect will be proportional to the electric-scalar potential. Thus, we demonstrated the suppression of the electric field while maintaining a time-varying electric potential within the output resonator. Using a signal generator locked to the central frequency of the field suppressed extreme dispersion characteristic, instead of a vector network analyser, would avoid instabilities, ensuring a precisely cancelled electric field throughout the duration of the experiment. Such precisely locked interferometers have been achieved with voltage-controlled devices for the development of low-phase noise oscillators \cite{ivanov1998microwave,ivanov2023frequency}, it may be necessary to operate the experiment in vacuum, with vibration isolation and temperature control, however theses are all standard experimental techniques that can be considered to realised a stable future experiment.

\begin{figure}[t]
\begin{minipage}{0.5\columnwidth}
\centering
\includegraphics[width=1\columnwidth]{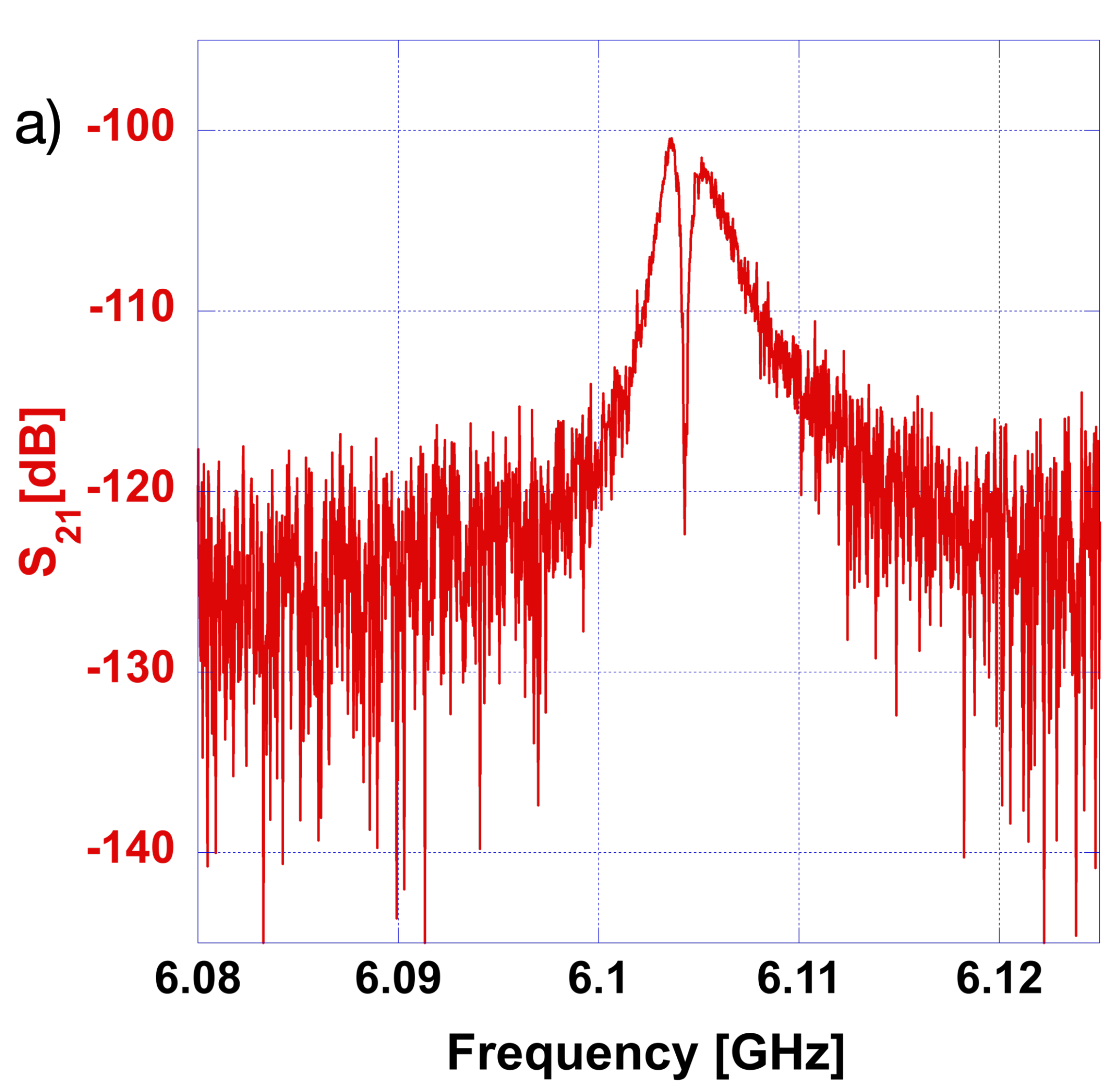}
\end{minipage}%
\begin{minipage}{0.5\columnwidth}
\includegraphics[width=1\columnwidth]{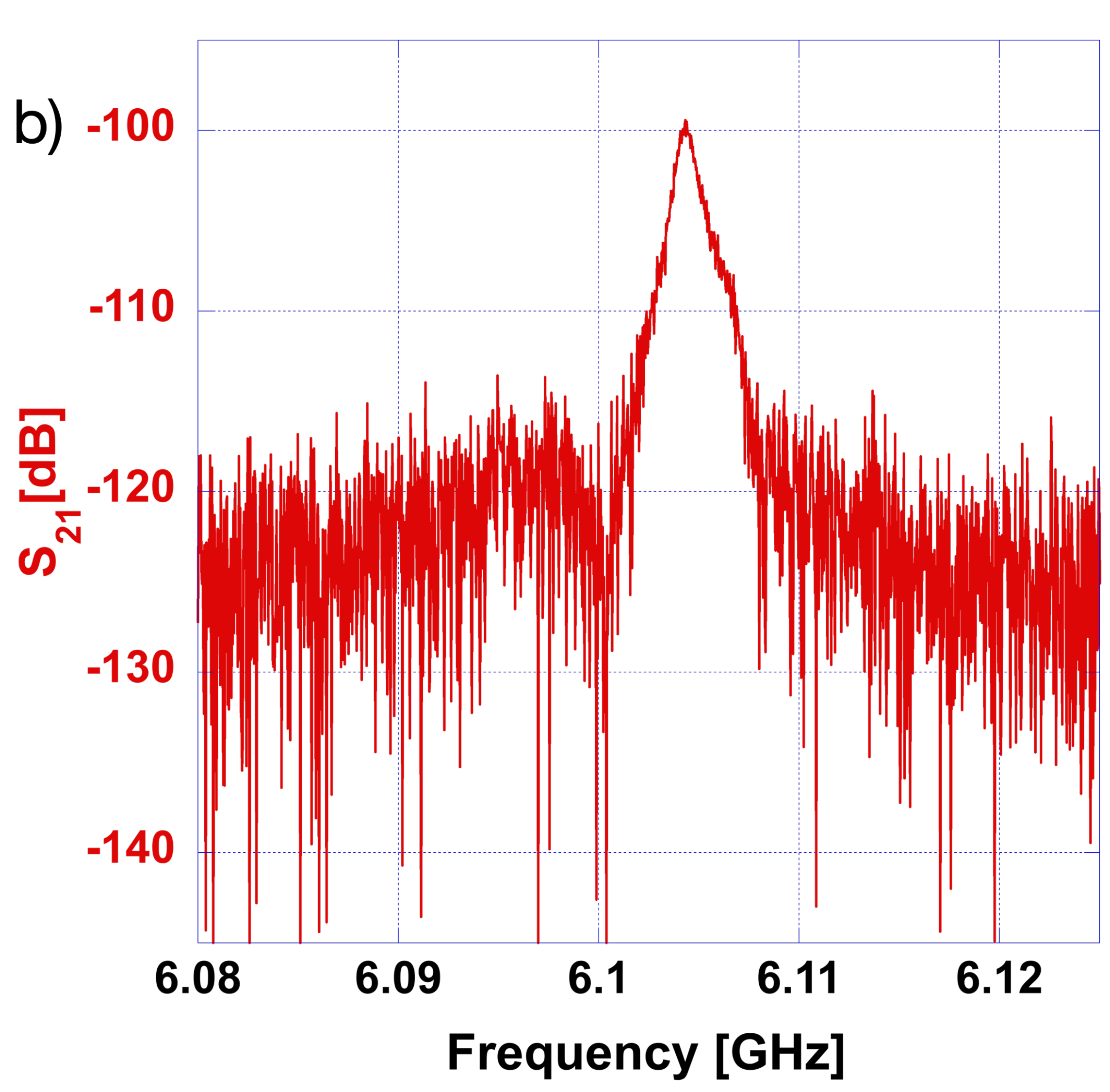}
\end{minipage}
\begin{minipage}{0.5\columnwidth}
\centering
\includegraphics[width=1\columnwidth]{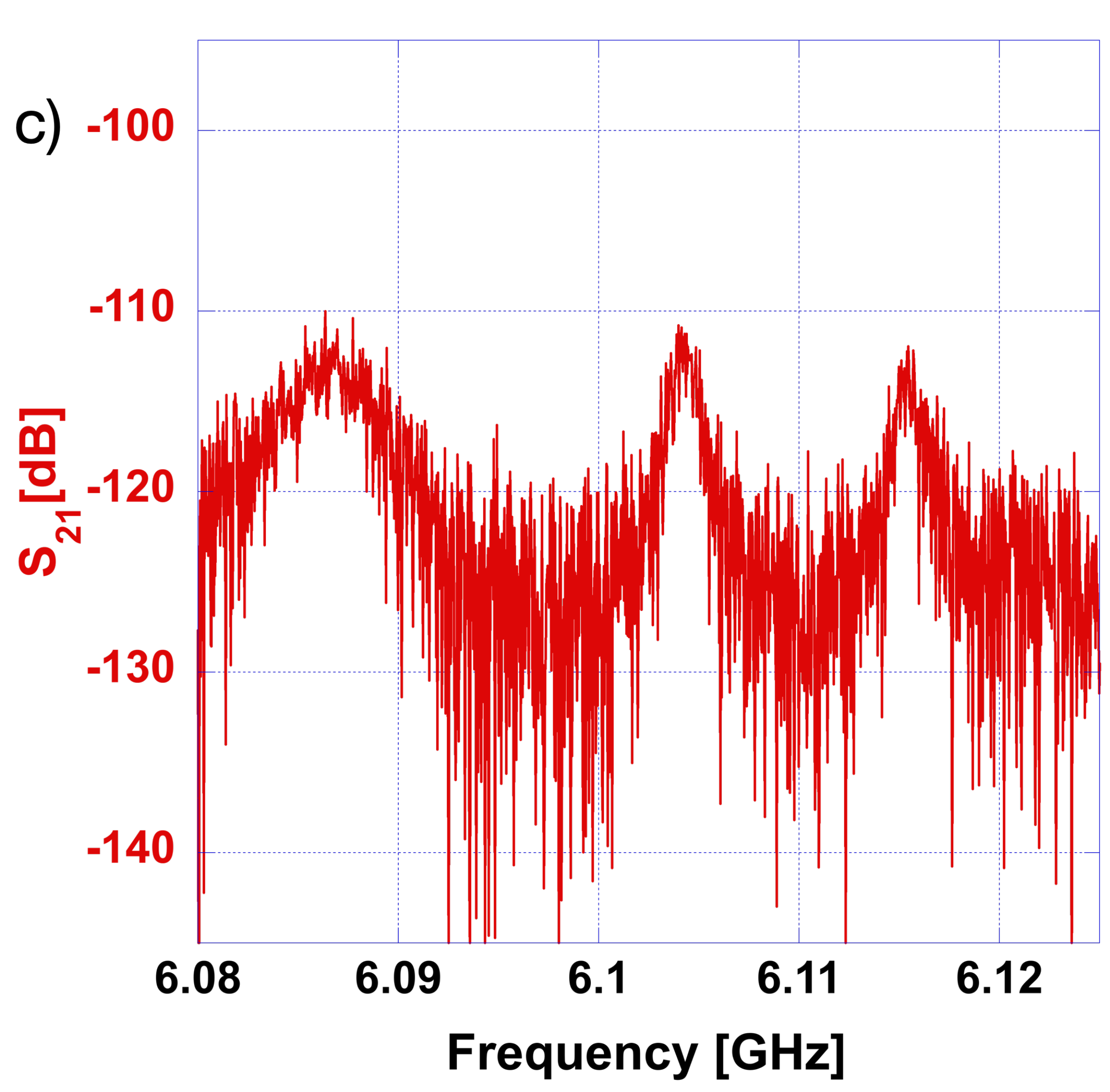}
\end{minipage}%
\begin{minipage}{0.5\columnwidth}
\raisebox{0pt}[\height][\depth]{\hspace{0.0cm}\includegraphics[width=0.97\columnwidth]{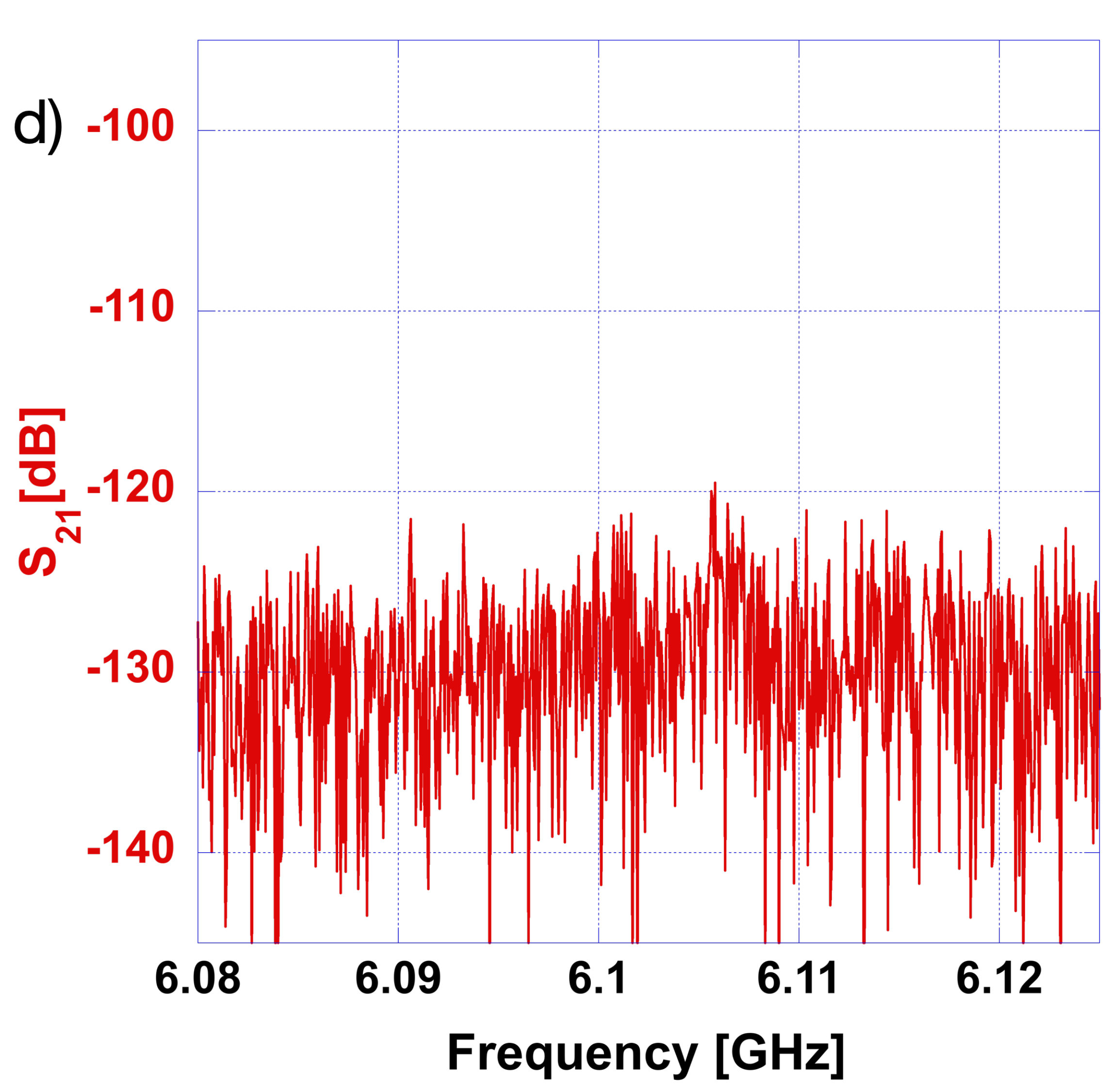}}
\end{minipage}
\caption{Some recorded $S_{21}$ transmission spectra as the input cavity frequencies are detuned from one another, starting from: (a) When all three frequencies are tuned and with the two input cavities in-phase in a similar state to that shown in Fig. \ref{phasetune}b. b) The input cavities are slightly detuned in frequency from the output cavity frequency, so the phase condition is no longer met, and the induced extreme dispersion effect disappears. (c) The application of a larger applied frequency detuning of the two input cavities so all three frequencies can be distinguished, also resulting in large attenuation of all three resonances. d) Transmission spectrum with a thick foil ($\approx$50 $\mu$m) as the intermediate between the input and output resonators instead of the thin ($<$8 $\mu$m) foil with all three resonant frequencies tuned. In this case the foils no longer transmit significant currents, charges and fields to excite photons in the output cavity mode irrespective of the phase and amplitude of the interferometer.}
\label{micrometertuning}
\end{figure}

\subsection{Further Verification}\!

To verify that the observed effect occurs only when the interferometer is balanced, a series of further tests were conducted. First, we confirmed that both input arms must be active to observe the field cancellation effect. To test this, the upper arm with the variable attenuator in Fig.~\ref{ABsetup} was deactivated by setting the attenuator to its maximum of 30 dB loss. In this scenario, transmission through the output resonance was observed from the central output cavity as it still could be driven from just one of the two foil boundaries. As expected, varying the phase of the interferometer had no effect on the transmission spectrum. Next, when both arms of the interferometer were activated, with a mismatch in the circulating power within the two cavities, variation in the transmission spectrum occurred, but with no observed anti-resonance conditions across the full range from the two phase shifters. These tests confirmed that not only must the phase of the input cavity modes be aligned, but the power must also be precisely matched to effectively annul the electric field. 

It was also evident that all three microwave cavity resonators must be tuned in frequency. Fig.~\ref{micrometertuning}a to c shows the effect of detuning the two input resonators away from the central output cavity frequency. Tuning the micrometers of the two input resonators to tune the frequency has the immediate effect of destroying the anti-resonance and increasing the central frequency transmission. 

Finally we implemented a much ticker coils of $\approx$50 $\mu$m in thickness, instead of the 8 $\mu$m foils. Due to the exponential skin-depth drop-off, there was no longer any significant current, charge or field penetration through the foil and hence photons in the output cavity could not be excited regardless of the level/matching of attenuation of the interferometer arms, nor the variation of the phase shifters while the input cavities are tuned to the output cavity resonance frequency.

\section{Conclusions}\!
In this work, we constructed a microwave cavity resonator, excited in an interferometric manner such that the electric field on resonance could be suppressed while maintaining a significant time varying electric-scalar potential. This phenomenon was shown to be equivalent to EIA, a form of extreme dispersion, which has led to an order of magnitude higher phase sensitivity. Because, the device has electric field suppression at the intended frequency, whilst maintaining a time-varying electric-scalar potential, the device should therefore have the capacity to probe the electric-scalar AB effect on an inserted quantum system. 

\section*{acknowledgements}
This work was funded by the ARC Centre of Excellence for Engineered Quantum Systems, CE170100009, and Dark Matter Particle Physics, CE200100008.

\end{document}